# Big Data: Understanding Big Data

Kevin Taylor-Sakyi
Engineering & Applied Science
Aston University
Birmingham, England
Kevin.sakyi@gmail.com

*Abstract*—Steve Jobs, one of the greatest visionaries of our time was quoted in 1996 saying *"a lot of times, people don't know what they want until you show it to them"*[38] **indicating he advocated products to be developed based on human intuition rather than research. With the advancements of mobile devices, social networks and the Internet of Things (IoT) enormous amounts of complex data, both structured & unstructured are being captured in hope to allow organizations to make better business decisions as data is now vital for an organizations success. These enormous amounts of data are referred to as <u>Big Data</u>, which enables a competitive advantage over rivals when processed and analyzed appropriately. However <u>Big Data Analytics</u> has a few concerns including** *Management of Data-lifecycle, Privacy & Security, and Data Representation.* **This paper reviews the fundamental concept of** *Big Data,* **the <u>Data Storage</u> domain, the <u>MapReduce programming paradigm</u> used in processing these large datasets, and focuses on two case studies showing the effectiveness of Big Data Analytics and presents how it could be of greater good in the future if handled appropriately.**

*Keywords—Big Data; Big Data Analytics; Big Data Inconsistencies; Data Storage; MapReduce; Knowledge-Space*

## I. Introduction

Frequently referred to as the information age, the economic industry in the 21$^{st}$ century is highly dependent on data. How much data must be processed to be of meaningful use? A study conducted by the IDC states only 0.5% of globally generated data is analyzed [3]. In a world where *"every 2 days we create as much information as we did from the beginning of time up to 2003"* [1] there is a need to bridge data analyzed with current trends to better business models; Systematic processing and analysis of big data is the underlining factor.

The purpose of this report is to reflect knowledge and understanding in the intriguing field of big data acquired through various papers from well-known fellows in the computing field as well as informational directories on the web. It aims to focus on the importance of understanding big data, envisioning the transformation from traditional analytics into big data analytics, data storage, and the future implications they'll have on business processes and big data in the years to come.

By 2020 there will be approximately 20-100 billion connected devices [2] leading to more data collection; thus illustrating a necessity for applying big data analytics. This brings forth the necessity of this survey report, understanding Big Data.

## II. Big Data

The phenomenon of *big data analytics* is continually growing as organizations remodel their operational processes to rely on live data with hope to drive effective marketing techniques, improve customer engagement, and to potentially provide new products and services [4] [13]. The questions to be raised are:

1) What is Big Data? (Section II)
2) Why is the transformation from traditional analytics to Big Data analytics necessary? (Section II)
3) How to meet demand for Computing Resources? (Section II)
4) What implications does Big Data have on the evolution of Data Storage? (Section III)
5) What are the inconsistencies of Big Data? (Section IV)
6) How is Big Data mapped into the knowledge space? (Section V)

### A. What is Big Data?

*Big Data* refers to large sets of complex data, both structured and unstructured which traditional processing techniques and/or algorithms are unable to operate on. It aims to reveal hidden patterns and has led to an evolution from a model-driven science paradigm into a data-driven science paradigm. According to a study by Boyd & Crawford [5] it *"rests on the interplay of:*

*(a) Technology: maximizing computation power and algorithmic accuracy to gather, analyze, link, and compare large data sets.*

*(b) Analysis: drawing on large data sets to identify patterns in order to make economic, social, technical, and legal claims.*

*(c) Mythology: the widespread belief that large data sets offer a higher form of intelligence and knowledge that can generate insights that were previously impossible, with the aura of truth, objectivity, and accuracy."*

IBM scientists mention that *Big Data* has four-dimensions: <u>Volume</u>, <u>Velocity</u>, <u>Variety</u>, and <u>Veracity</u> [7]. Gartner agrees with IBM by stating that *Big Data* consists of *"high-volume,*

*velocity and variety information assets that demand cost-effective, innovative forms of information processing for enhanced insight and decision making*" [8].

Description of the four-dimensions are detailed below [39]:

- Volume – Current data existing is in petabytes, which is already problematic; it's predicted that in the next few years it's to increase to zettabytes (ZB) [39]. This is due to increase use of mobile devices and social networks primarily.
- Velocity – Refers to both the rate at which data is captured and the rate of data flow. Increased dependability on live data cause challenges for traditional analytics as the data is too large and continuously in motion.
- Variety – As data collected is not of a specific category or from a single source, there are numerous raw data formats, obtained from the web, texts, sensors, e-mails, etc. which are structured or unstructured. This large amount causes old traditional analytical methods to fail in managing big data.
- Veracity – Ambiguity within data is the primary focus in this dimension – typically from noise and abnormalities within the data.

Equipping an enterprise with *Big Data* driven e-commerce architecture aids in gaining extensive "*insight into customer behaviour, industry trends, more accurate decisions to improve just about every aspect of the business, from marketing and advertising, to merchandising, operations, and even customer retention*"[9]. However the enormous data obtained may challenge the Four V's mentioned earlier. Referring to the following examples, insights into how Big Data brings value to organizations are mentioned.

United Parcel Service (UPS) grasped the importance of big data analytics early in 2009 leading to the installation of sensors on over 46,000 of its vehicles; the idea behind this was to attain "*the truck's speed and location, the number of times it's placed in reverse and whether the driver's seat belt is buckled. Much of the information is uploaded at the end of the day to a UPS data center and analyzed overnight*" [10]. By analyzing the fuel-efficiency sensors and the GPS data, UPS was able to reduce the consumption of fuel by 8.4 million gallons and reduced the duration of its routes at 85 million miles [10].

Andrew Pole, data analyst for American retailer Target Corporation, developed a pregnancy-predictive model, as indicated by the name customers are assigned a "pregnancy prediction" [11] score. Furthermore Target was able to gain insight into how 'pregnant a woman is'. Pole initially used Target's *baby-shower-registry* to attain insight into women's shopping habits, discerning those habits as they approached their due date; which "*women on the registry had willingly disclosed*" [11]. Following the initial data collection phase, Pole and his team ran series of tests, analyzed the data sets and effectively concluded to patterns that could be of use to the corporation.

Within the model each customer is assigned a unique number, internally classified as a *Guest ID number*. Linked with this is a shoppers purchased products, methods of previous payments (*coupons, cash, credit cards, etc.*), virtual interaction (*clicking links in sent e-mails, customer service chat, etc.*). Compiling these data sets along with demographic data that is available for purchase from information service providers such as *Experian* [12] and the alike, Target is able to channel their marketing strategy effectively. In this case, grasp a pregnant shopper's attention by sending a coupon via e-mail or post, which can be distinguished by analyzing previous effective methods. Additionally, by obtaining shoppers demographic data, Target is able to trigger a shopper's habit by including other products that may not commonly be purchased at Target, *i.e. milk, food, toys, etc.* Contributing to a valuable competitive advantage over competitors during the early years when the model was unrevealed to the public.

Big Data should not be looked merely as a new ideology but rather as a new environment, one that requires "*new understanding of data collection, new vision for IT specialist skills, new approaches to security issues, and new understanding of efficiency in any spheres*" [27]. This environment, when analyzed and processed properly enhances business opportunities; however the risks involved should be taken into account when collecting, storing and processing these large data sets.

### B. Big Data analytics transformation

In assessing the grounds on why several organizations are gravitating towards *Big Data* analytics, concrete understanding of traditional analytics is necessary. Traditional analytical methods include structured data sets that are periodically queried for specific purposes [6]. A common data model used to manage and process commercial applications is the relational model; this Relational Database Management Systems (RDBMS) provides "*user-friendly query languages*" and provides simplicity that other network or hierarchical models are not able to deliver. Within this system are tables, each with a unique name, where related data is stored in rows and columns [28].

These streams of data are obtained through integrated databases and provide leverage on the intended use of those data sets. They do not provide *much* advantage for the purpose of creating newer products and/or services as *Big Data* does – leading to the transformation of *Big Data* analytics.

Frequent usage of mobile devices, Web 2.0, and growth in the Internet of Things are among a few to mention in reasoning behind organizations looking to transform analytical processes. Organizations are attracted to big data analytics as it provides a means of obtaining real time data, suitable for enhancing business operations. Along with providing parallel & distributed processing architectures in data processing, big data analytics also enables the following services: "*after-sales service, searching for missing people, smart traffic control system, customer behavior analytics, and crisis management system*" [34].

Research conducted by Kyounghyun Park and his colleagues of the *Big Data SW Research Department* in South Korea proved to develop a platform with the purpose of establishing <u>Big Data as a Service</u>. In facilitating the services mentioned above, traditional methods require various distinctive systems to perform various tasks such as "*data collection, data pre-processing, information extraction, and visualization*" [34]. On the contrary, this web-based platform provides developers & data scientist an environment to "*develop cloud services more efficiently*", view shared data, "*supports collaborative environments to users so that they can reuse other's data and algorithms and concentrate on their own work such as developing algorithms or services*" [34]. Unlike traditional methods where databases are accessible by all persons, this platform and similar big data analytic platforms supports restricted access on different datasets.

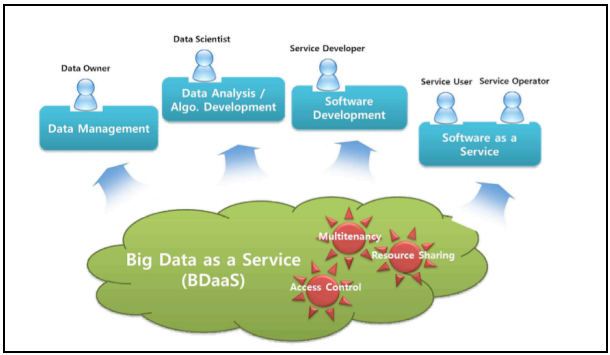

Fig. 1. Collaborative Big Data platform concept for Big Data as a Service[34]

This big data platform and the like prove their cost efficiency by:

- Centralizing all aspects of storage and processing procedures in big data onto one platform
- Ensures easy and rapid access to view other data; allowing developers to focus on their own work (algorithms/service)
- Privacy coverage

Currently there are over 4.6 billion mobile phone subscribers. In addition, there are over 1 to 2 billion persons accessing the Internet at any given time [1]. Two of the most widely known social platforms:

- <u>Facebook</u> has over 1 billion active users monthly, accumulating over 30 billion bits of shared content.
- <u>Twitter</u> also has mass amounts of data, serving as a platform for over 175 million *tweets* a day [1].

With approximately 2.5 quintillion bytes of data being created everyday [28], it is understandable why *Business Intelligence* (BI) is drifting towards analytical processes that involve extraction of larger data sets, as traditional management systems are unable to fathom these amounts of data. "*Data are being constantly collected through specially designed devices to help explore various complex systems*" [4]. These streams of data when analyzed properly using big data methods will "*help predict the possibility to increase productivity, quality and flexibility*" [4]. The power of big data is its ability to bring forth much more intelligent measures of formulating decisions.

*C. MapReduce*

The advancements in technology within the last few decades has caused an explosion of data set sizes [29], though there's now <u>more to work with</u> the speed at which these volumes of data are growing exceeds the computing resources available. **MapReduce**, a programming paradigm utilized for "*processing large data sets in distributed environments*" is looked upon as an approach to handle the increased demand for computing resources [29].

There are two fundamental functions within the paradigm, the <u>Map</u> and the <u>Reduce function</u>. The **Map function** executes sorting and filtering, effectively converting a data set into another & the **Reduce function** takes the output from the *Map* function as an input, then completes grouping and aggregation operations [29] to combine those data sets into smaller sets of tuples [30]. *Figures 1 & 2* represent the computational formulas to perform *MapReduce*.

```
function map(name, document)
  for each word in document
    emit (word, 1)
```

Fig. 2. Map function

The *Map* function above effectively takes a document as an input, split the words contained in a document (*i.e. a text file)*, and creates a (Key, Value) pair for each word in the document.

```
function reduce (word, List partialCounts)
  sum = 0
  for each pc in partialCounts
    sum += pc
  emit (word, sum)
```

Fig. 3. Reduce function

In the *Reduce* function the list of Values (*partialCounts*) are worked on per each Key (*word*). "*To calculate the occurrence of each word, the Reduce function groups by word and sums the values received in the partialCounts list*" [29].

As the final output, a list of words displaying their occurrences in the document is displayed. The *MapReduce paradigm*'s main perk is its scalability, it "*allows for highly parallelized and distributed execution over a large number of nodes*". An open source implementation of *MapReduce* is *Hadoop* [29]. Within this implementation, tasks in the *Map* or *Reduce* function are separated into various *jobs*. These *jobs* are assigned to nodes in the network, and assigned to other nodes if an initial node fails its *jobs*. Hadoop permits the distribution of big data processes across machines using simple programming models not suitable for big data [4].

Though *MapReduce* provides significant progress in data

storage, a few uncertainties persist within this paradigm: [29]

1) Absence of a *standardized SQL query language*:
   i. Current solution, providing SQL on-top of MapReduce [29]: Apache Hive - stipulates an SQL-like language on top of Hadoop [31]
   ii. Deficiency in data management features like advanced indexing and a complex optimizer [29]
   iii. NoSQL Solutions - MongoDB & Cassandra – enable queries similar to SQL HBase uses Hive [32]

2) "*Limited optimization of MapReduce jobs*" [29]
   i. "*Integration among MapReduce, distributed file system, RDBMSs and NoSQL stores*" [29]

### III. DATA STORAGE

Data storage has always been an area of concern in the knowledge management domain. Shortly after the *Information Explosion* was conceptualized [15] in the 1930s, focus was driven to understand how to manage this everlasting growth of data and information. Referring to libraries, the first source of data organization and storage, a sign of data storage overload was reported in 1944 when Fremont Rider, a librarian from Wesleyan University, estimated that "*American University libraries were doubling in size every 16 years*" [14].

With this estimation, it was necessary to change the methods of storing and retrieving data, not only in respect to libraries but all sectors concerned with knowledge management. Before analyzing the importance of effective data storage, a better understanding in the evolution of data storage is briefly documented:

*A. Evolution of Data Storage*

- Late *1920s* – IBM takes successfully redesigns Basile Bouchon's punch card invention, generating 20% of their revenue in the 1950s [17]
- 1952 – IBM announces first magnetic tape storage unit; standard data storage technology in the 1950s and still in use in the entertainment industry [16]
- 1956 – IBM invents first hard drive first hard drive capable of holding up to 5MB, pushing to 1GB in 1982, and a few TB currently
- 1967 – First floppy disk is created, initially storing up to 360KB on a 5.25-inch disk leading to a 3.5-inch disk capable of storing 1.44MB [18]
- 1982 - Concept of the compact disk (CD) is invented in Japan, CD-ROM is later developed with storage capacity of 650MB to 700MB; Equivalent to 450 floppy disks [19]
- 2000 – USB flash drives debuts; similar to the floppy disks, data storage capacity improved overtime and continually improves [20]
- 2010s – "*The Cloud is estimated to contribute more than 1 Exabyte of data*" [20]

In the early 1950s Fritz-Rudolph Güntsch developed the concept of *virtual memory*, allowing finite storage to be treated as infinite [14]. Güntsch's concept permitted the processing of "*data without the hardware memory constraints that previously forced the problem to be partitioned*"[14] in other words, focus was on the hardware architecture not the data itself. Derek Price, information scientist known as the *father of scientometrics*, [21] agreed with Rider's statement on storage overload by observing and expressing that "*the vast amount of scientific research was too much for humans to keep abreast of*" [14]. The evolution of data storage has shown to continually improve techniques of storage, effectively storing '*more on less*', now dealing away with most in-house hardware's and focusing on cloud storage. However this method of mass data storage may be problematic if not prepared for accordingly.

Data storage is now extremely complex. Big data or '*total data landscape*' as 451 Researchers call it [22] reflect complications not only with storage techniques, but also with the transformation of those data-sets to map them into *knowledge-space* in order to drive some value to a business as well as the costs involved.

As Professor Hai Zhuge of Aston University mentioned, "*If a problem is unclear or unable to be represented, problem-solving is not meaningful*" [4]. This is a fundamental issue with big data analytics. Viewing the *Evolution of Data Storage* documented earlier it's evident that engineers have always been eager to compress more data into smaller storage units, and now into the cloud.

Data capture has escalated from manual inputs in the early years of technology to now attaining data sets from social media platforms, IoT devices, and the like. Availability for the '*general population*' to create data through their mobile devices and computers strike as an opportunity for businesses to improve their business models, the storage mediums earlier available in the early 2000s were problematic. Modern technology now allows approximately 4 terabyte ($10^2$) per disk, equating to 25,000 disks for 1 terabyte ($10^{18}$) [29].

Some view this as being problematic in the life science space as they involve large datasets; Looking to the example below a systematic analysis will be shown as to why this point of view is not accurate.

Example scenario: *Scientific research in the field of the Human Genome.*

In the domain of *human genome* research, an abundance of data is collected. There are about 2 terabytes in a single operation when obtaining "*a single sequence of one individual human*" [22]; to grasp a wider understanding this equates to 2000hrs or 80 days of videos. However when conducting research of this magnitude 2 terabytes is merely a starting point, a data scientist in 2012 stated 2 terabytes can undoubtedly push to nearly 7 terabytes once processing and analysis commences [16]. This large amount of data represents a single human sequence, and in confirming results, scientists tend to attain more data – requiring more capacity for storage.

Rather than executing various experiments individually in hope to obtain information, researchers grasped the opportunity of establishing a method in obtaining more concrete data from various locations. In the early years of the Human Genome Project (HGP) (1993) researchers realized it was imperative to "*assess the contribution of informatics to the ultimate success of the project*" [24]. This led to the implementation of three different databases used throughout the HGP project, [24] prompting a linkage between the primary databases. As an initial proposal, the following was recommended: "*a link between the sequence and mapping database can be made by using the common sequence identity between the sequence record and sequences stored in the mapping database*" [24]. Though a reasonable proposal, a key requirement for the method to be effective required the following: constant, informative and no overlapping of classification. With the possibility of triggering chaos, protocols were established, *i.e. 'establishing naming conventions'* as several genes were cloned [24] – complimenting Zhuge's viewpoint on problems having to be transparent in order to be comprehendible.

In all, due to the magnitude of data stored, processed and analyzed for the HGP, about $2.7 billion of U.S. taxpayers was used in FY 1991 to complete the project; about $4.74 billion in comparison to the 2015 FY. Sequencing a human genome is now possible at an exceedingly low price of $1000 in a matter of days – outperforming the core of *Moore's Law*.

The evolution of *the cloud* has enabled this extreme change of cost. *Infrastructure as a Service* (IaaS) providers such as Amazon's S3 storage system & Elastic Compute cloud (EC2) provides scalable services to store any amount of data & targets organizations to utilize their idle asset at hourly rate. These infrastructures were originally developed to upkeep its retail businesses, yet because they altered their strategy Amazon is now the most high-profiled IaaS service providers. Their Web Services currently holds about 27% of all cloud IaaS with data centres in 190 countries [25]; because of this innovative concept of shifting from a physical web server to cloud architecture in 2006, other services catered to specific data storage & management have emerged [23].

Several platforms are now available to aid in the accessibility of data and storage issue raised within the *HGP*. One of these platforms, *DNAnexus,* "*provides a global network for sharing and management of genomic data and tools to accelerate genomic medicine*" [26]. *DNAnexus's* web interface offers easy automatic uploads to scalable infrastructure of data and "*analysis & visualization tools for DNA sequencing*" [25]. This platform allows researchers and organizations to bypass the complication of dealing with protocols, naming conventions, and the like; empowering researchers to focus on actual data interpretation that may be used to further develop predictive medical treatments rather than manual processing and/or analysis.

Effective data storage systems are those that classify the retrieved information in an appropriate format for value extraction and analysis. Researchers in China believe for this to occur the subsystem must provide the following [35]:

- "Storage infrastructure must accommodate information persistently and reliably"
- "A scalable access interface to query and analyze a vast quantity of data"

IV. BIG DATA INCONSISTENCIES

With the emergence of the *Information Era*, computational ability to capture data has increased considerably; sources of these data streams come from all sectors: automobile, medical, IT, and more. Majority of data used in *Big Data* analytics come in unstructured formats, primarily obtained from the Internet (*cloud computing*) and social media. According to industry consultant leaders at *CapGemini*, organizations looking to optimize their big data processes need to refer to *tweets, text messages, blogs,* and the like to recognize response on specific products or services that may aid in discovering new trends [33]. *Web 2.0* and the *Internet of Things* are among the main sources by which data capture is enabled. Management of data life cycle, data representation, and data privacy & security are among the pressing issues within big data analytics.

A. Management of Data life-cycle

Pervasive sensing through *tablets, smart phones, and sensor-based-Internet-devices* contributes towards the exceptional rates and scale at which data is being obtained, surpassing the current storage technologies available. The usefulness of big data analytics is to reveal value from fresh data; however, the colossal amounts of data being captured pose an issue for discovering 'useful' information at specific times.

In preventing or limiting the amount of '*value loss*' within these systems principles should be governed, outlining and determining when and which datasets should be either archived for further usage or discarded [35].

B. Data Privacy & Security

Privacy & security are considered to be the most important issue within big data analytics [36]. As organizations look to utilize data to provide personalized information for the strategy of their operations, *legal ramifications & context-based* are of high concerns.

In the field of medical research several scientists are to share wide streams of data to gather in finding solutions of creating new medications or diagnosing new diseases. It's believed that by 2020 there will be approximately 12 ZBs of data in the cloud from *Electronic Healthcare Records, Electronic Medical Records, Personal Health Records,*

*Mobilized Health Records, and mobile monitors* alone [37]. The growths of these datasets are in line with the increasing importance that they have accumulated. Conserving data privacy is an increasing issue because of the vast amount of data captured; though there are some guidelines & laws to keep these various data sources secure there are flaws within them [36]. With different countries having different policies, a new technique considering a uniform *global* aspect must be established to insure that all data on the cloud are constrained equally among the different locations where they're accessed. *(I.e. a scientist accessing data in England from a source in China should have the same privacy as in China.)*

The plethora of context based information (*i.e. linking personal data of a person's social media information with other attained data to distinguish new information*) for analytics instigates complexity in defining which sets of data are to be classified as sensitive as all sets have different meaning in different context. As a result employing privacy on such data sets become difficult.

### C. Data Representation

Data representation is the means of representing information stored on a computer from different types of data: *texts, numerical, sound, graphical (video & images)*, etc. In analytics, not only do the datasets come in different types, their "*semantics, organization, granularity, and means of accessibility*" are also diverse [35]. Zhuge mentions computers are difficult to determine the correctness of representation, referring to the concepts of calculating meaning and explaining the results obtained [4].

H. Hu and his colleagues, fellows of the *IEEE* proposed the following as a measure to avoid representation issues:

- Presentation of data must be designed not to merely display singularity of data but rather reflect the "*structure, hierarchy, and diversity of the data, and an integration technique should be designed to enable efficient operations across different datasets*" [35]

### V. MAPPING INTO KNOWLEDGE-SPACE

After data is captured, processed and analyzed, how are they mapped into other dimensions? Dimensions within the big data environment are methods for *observing, classifying, and understanding a space* [4]. Knowledge space acquires the structure of a semantic network where its vertices represent knowledge modules and its relations represent relations amongst two knowledge modules [40].

A multi-dimensional resource space model was proposed to manage knowledge from multiple dimensions. It provides a way to divide big data into a multi-dimensional resource space through multi-dimensional classifications [41][42].

To bridge the gap between machines and humans in order to map data into knowledge space appropriately, a cyber-space infrastructure must be formulated. The United States defines this infrastructure as "*the converged information technologies including the Internet, hardware and software that support a technical platform where users can use globally distributed digital resources*" [4]. This infrastructure provides cross-connections in the space - interlocking different modules within the different dimensions.

Representation of knowledge from big data analytics requires multiple links throughout various spaces (*i.e. physical, social, knowledge, etc.*) to not only link different symbols but also to differing individuals [41]. Vannevar Bush's establishment of the theoretical *Memex* machine awakened further study and research into interlinking various spaces through the *cyber-space* [41].

Within the cyber-space, continuous tests must be conducted to ensure the capability of modules being derived from other modules, no conflicts between modules, and no partial modules as this could lead to ambiguous results.

A number of softwares have been developed to accommodate the issues brought forth by establishing a well-formulated cognitive cyber-infrastructure. *Globus,* for example, has been developed to accommodate the integration of "*higher-level services that enable applications to adapt to heterogeneous and dynamically changing meta-computing environments*" [4].

Knowledge-space has strong impacts to scientific conceptions that may deliver tools to be used in identifying strong points. Value of statements can be tested as a result of the knowledge-space [40]. Intelligent manufacturing is a result of such space, humans mapping the representation of the external world into their mind (*i.e. social & economical factors not computable*), converging it onto the virtual space, and allowing the different modules to be interlocked, displaying weak-points and strong-points, bringing forth the $4^{th}$ industrial revolution.

### VI. CONCLUSION

The concept of Big Data analytics is continually growing. Its environment demonstrates great opportunities for organizations within various sectors to compete with a competitive advantage, as shown in the examples mentioned earlier. The future of medical science is changing dramatically due to this concept, scientist are able to access data rapidly on a global scale via the cloud, and these analytics contribute to the development of predictive analytic tools (*i.e. facilitating predictive results at primary stages*). However as mentioned (*Section IV*), there are inconsistencies and challenges within *Big Data privacy: sufficient encryption algorithms to conceal raw data or analysis, reliability & integrity of Big Data, data storage issues and flaws within the MapReduce paradigm.* This paper shed light on conceptual ideologies about *big data analytics* and displayed through a few scenarios how it's beneficial for organizations within various sectors if analytics are conducted correctly. Further areas to research to grasp a wider understanding of Big Data are data processing & data transfer techniques.